\documentclass[longauth,structabstract]{aa} 
\usepackage{graphicx}
\usepackage{natbib}
\usepackage{txfonts}
\begin{document}
   \title{Reversal of infall in SgrB2(M) revealed by \textit{Herschel}\thanks{\textit{Herschel} is an ESA space
    observatory with science instruments provided by European-led
    Principal Investigator consortia and with important participation
    from NASA.}/HIFI observations of HCN lines at THz frequencies}


\author{
R.~Rolffs,\inst{1,2}
P.~Schilke,\inst{1,2}
C.~Comito,\inst{1}
E.~A.~Bergin,\inst{3}
F.~F.~S.~van der Tak,\inst{4}
D.~C.~Lis,\inst{5}
S.-L.~Qin,\inst{2}
K.~M.~Menten,\inst{1}
R.~G\"usten, \inst{1}
T.~A.~Bell,\inst{5}
G.A. Blake,\inst{6}
E.~Caux,\inst{7,8}
C.~Ceccarelli,\inst{9}
J.~Cernicharo,\inst{10}
N.~R.~Crockett,\inst{3}
F.~Daniel,\inst{10,11}
M.-L.~Dubernet,\inst{12,13}
M.~Emprechtinger,\inst{5}
P.~Encrenaz,\inst{11}
M.~Gerin,\inst{11}
T.~F.~Giesen,\inst{2}
J.~R.~Goicoechea,\inst{10}
P.~F.~Goldsmith,\inst{14}
H.~Gupta,\inst{14}
E.~Herbst,\inst{15}
C.~Joblin,\inst{7,8}
D.~Johnstone,\inst{16}
W.~D. Langer,\inst{14}
W.D.~Latter,\inst{17} 
S.~D.~Lord,\inst{17}
S.~Maret,\inst{9}
P.~G.~Martin,\inst{18}
G.~J.~Melnick,\inst{19}
P.~Morris,\inst{17}
H.~S.~P. M\"uller,\inst{2}
J.~A.~Murphy,\inst{20}
V.~Ossenkopf,\inst{2,4}
J.~C.~Pearson,\inst{14}
M.~P\'erault,\inst{11}
T.~G.~Phillips,\inst{5}
R.~Plume,\inst{21}
S.~Schlemmer,\inst{2}
J.~Stutzki,\inst{2}
N.~Trappe,\inst{20}
C.~Vastel,\inst{7,8}
S.~Wang,\inst{3}
H.~W.~Yorke,\inst{14}
S.~Yu,\inst{14}
J.~Zmuidzinas,\inst{5}
M.C.~Diez-Gonzalez,\inst{22}	
R.~Bachiller,\inst{22}	
J.~Martin-Pintado,\inst{23}
W.~Baechtold,\inst{24}	
M.~Olberg,\inst{25,4}	
L.H.~Nordh,\inst{26}
J.J.~Gill\inst{14}
\and
G.~Chattopadhyay\inst{14}
}
\institute{Max-Planck-Institut f\"ur Radioastronomie, Auf dem H\"ugel 69, 53121 Bonn, Germany\\
                \email{rrolffs@mpifr.de}
\and I. Physikalisches Institut, Universit\"at zu K\"oln,
              Z\"ulpicher Str. 77, 50937 K\"oln, Germany
\and Department of Astronomy, University of Michigan, 500 Church Street, Ann Arbor, MI 48109, USA 
\and SRON Netherlands Institute for Space Research, PO Box 800, 9700 AV, Groningen, The Netherlands
\and California Institute of Technology, Cahill Center for Astronomy and Astrophysics 301-17, Pasadena, CA 91125 USA
\and California Institute of Technology, Division of Geological and Planetary Sciences, MS 150-21, Pasadena, CA 91125, USA
\and Centre d'\'etude Spatiale des Rayonnements, Universit\'e de Toulouse [UPS], 31062 Toulouse Cedex 9, France
\and CNRS/INSU, UMR 5187, 9 avenue du Colonel Roche, 31028 Toulouse Cedex 4, France
\and Laboratoire d'Astrophysique de l'Observatoire de Grenoble, 
BP 53, 38041 Grenoble, Cedex 9, France.
\and Centro de Astrobiolog\'ia (CSIC/INTA), Laboratorio de Astrof\'isica Molecular, Ctra. de Torrej\'on a Ajalvir km 4,
28850 Madrid, Spain
\and LERMA, CNRS UMR8112, Observatoire de Paris and \'Ecole Normale Sup\'erieure, 24 Rue Lhomond, 75231 Paris Cedex 05, France
\and LPMAA, UMR7092, Universit\'e Pierre et Marie Curie,  Paris, France
\and  LUTH, UMR8102, Observatoire de Paris, Meudon, France
\and Jet Propulsion Laboratory,  Caltech, Pasadena, CA 91109, USA
\and Departments of Physics, Astronomy and Chemistry, Ohio State University, Columbus, OH 43210, USA
\and National Research Council Canada, Herzberg Institute of Astrophysics, 5071 West Saanich Road, Victoria, BC V9E 2E7, Canada 
\and Infrared Processing and Analysis Center, California Institute of Technology, MS 100-22, Pasadena, CA 91125
\and Canadian Institute for Theoretical Astrophysics, University of Toronto, 60 St George St, Toronto, ON M5S 3H8, Canada
\and Harvard-Smithsonian Center for Astrophysics, 60 Garden Street, Cambridge MA 02138, USA
\and  National University of Ireland Maynooth, Ireland
\and Department of Physics and Astronomy, University of Calgary, 2500 University Drive NW, Calgary, AB T2N 1N4, Canada
\and Observatorio Astron\'omico Nacional (IGN), Centro Astron\'omico de Yebes, Apartado 148. 19080 Guadalajara,  Spain
\and Departamento de Astrofı\'isica Molecular e Infrarroja, Instituto de Estructura de la Materia, CSIC, Calle Serrano 121, 28006 Madrid, Spain
\and Microwave Laboratory, ETH Zurich, 8092 Zurich, Switzerland
\and	Chalmers University of Technology, SE-412 96 G\"oteborg, Sweden, Sweden
\and	Department of Astronomy, Stockholm University, SE-106 91 Stockholm, Sweden
}


   \date{Received ?}

 
  \abstract
   {} 
   {To investigate the accretion and feedback processes in massive star formation, we analyze the shapes of emission lines from hot molecular cores, whose asymmetries trace infall and expansion motions.}
   {The high-mass star forming region SgrB2(M) was observed with \textit{Herschel}/HIFI (HEXOS key project) in various lines of HCN and its isotopologues, complemented by APEX data. The observations are compared to spherically symmetric, centrally heated models with density power-law gradient and different velocity fields (infall or infall+expansion), using the radiative transfer code RATRAN. }
   {The HCN line profiles are asymmetric, with the emission peak shifting from blue to red with increasing J and decreasing line opacity (HCN to H$^{13}$CN). This is most evident in the HCN 12--11 line at 1062 GHz. These line shapes are reproduced by a model whose velocity field changes from infall in the outer part to expansion in the inner part.}
   {The qualitative reproduction of the HCN lines suggests that infall dominates in the colder, outer regions, but expansion dominates in the warmer, inner regions. We are thus witnessing the onset of feedback in massive star formation, starting to reverse the infall and  finally disrupting the whole molecular cloud. To obtain our result, the THz lines uniquely covered by HIFI were critically important.}

\titlerunning{Reversal of infall in SgrB2(M)}
 \authorrunning{Rolffs et al.}
  \maketitle
%

\section{Introduction}

Massive stars form deep inside molecular clouds. While this obscures the process at infrared or shorter wavelengths, it offers the possibility of investigating high-mass star formation using dust emission and molecular lines from the radio to the far-infrared. The shapes of molecular lines are a particularly powerful tool, since they trace the density and temperature gradients as well as the velocity field. Even when the cores are not spatially resolved, radiative transfer modeling of a variety of lines can constrain their structure. The unique spectral coverage of HIFI opens up the full potential of this technique, since both the excitation of rotational transitions and the dust opacity increase with frequency. While the former allows us to study higher temperatures and densities, the latter leads to stronger continuum radiation, thereby facilitating the production of absorption features.

A particularly important stage of high-mass star formation is the hot core phase. When massive stars begin burning hydrogen, they are still accreting and surrounded by dense molecular gas, which is heated and emits a rich spectrum of molecular lines. Various feedback mechanisms (radiation pressure, pressure increase due to ionization, stellar winds, bipolar outflows)  push the gas outwards. Finally, this dominates over gravitational accretion, and the molecular cloud is disrupted.

The giant molecular cloud Sagittarius B2 is the most massive and active star-forming region in our Galaxy. It is located close to the Galactic center at a distance of 7.8 kpc \citep{Reid09}. SgrB2(Main) 
is a tight cluster of ultracompact H{\sc ii} regions \citep{Gaume90}, associated with a massive hot molecular core. Its luminosity is estimated to be $6.3\times10^6$ L$_\odot$ \citep{Goldsmith92}, the most luminous hot core of the Galaxy. Both infall and a decelerating outflow were seen by \citet{Qin08}.

In this paper, we present new results for the velocity field of SgrB2(M), based on \textit{Herschel}/HIFI observations of HCN and complemented by APEX data.

\begin{figure*}
  \centering
  \includegraphics[angle=-90,width=0.9\textwidth]{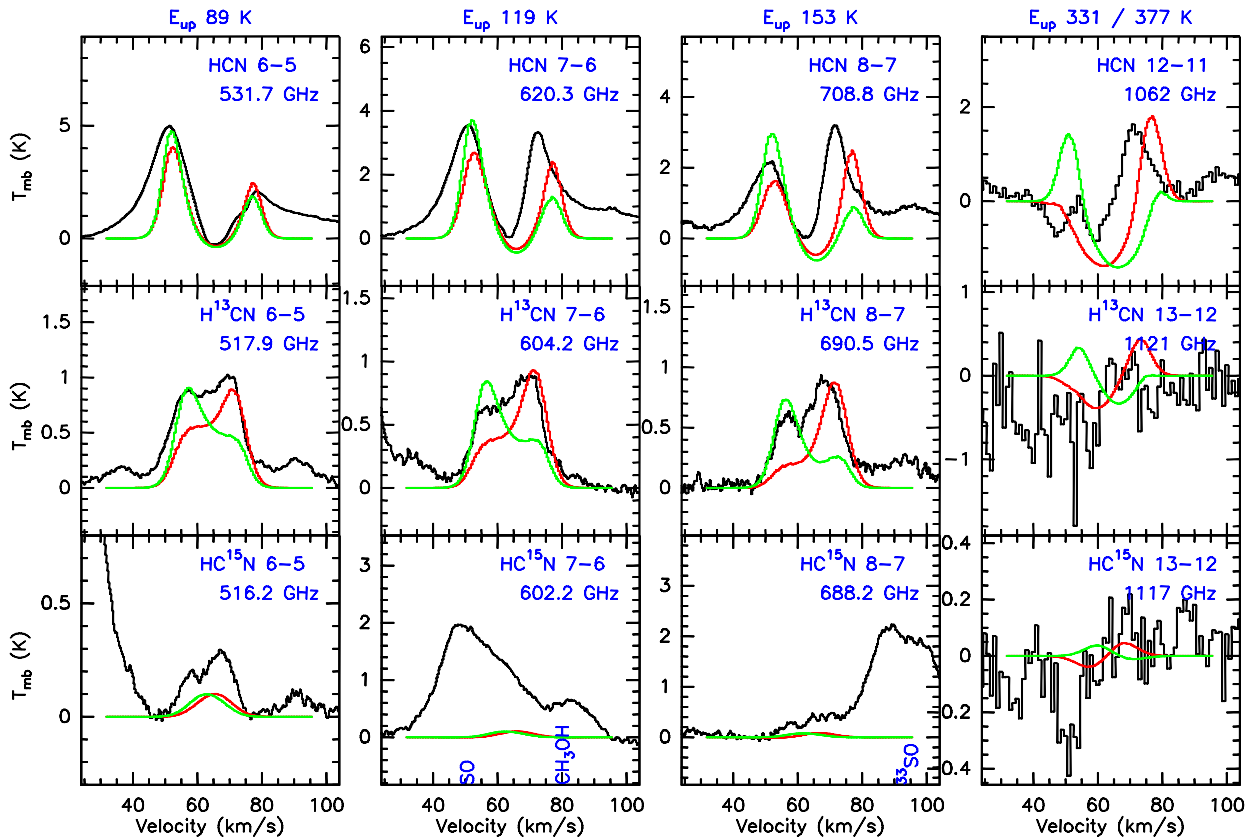}
  \caption{\textit{Herschel}/HIFI observations of HCN in SgrB2(M). The green model has a constant infall, the red model has infall in the outer part and expansion in the inner region. Note the different  asymmetries in the lines, which are reproduced by the red model. Upper state energies are stated above the plot.}
  \label{fig:hifi_hcn}
\end{figure*}

\section{Observations}

\begin{figure}
  \centering
  \includegraphics[angle=-90,width=0.49\textwidth]{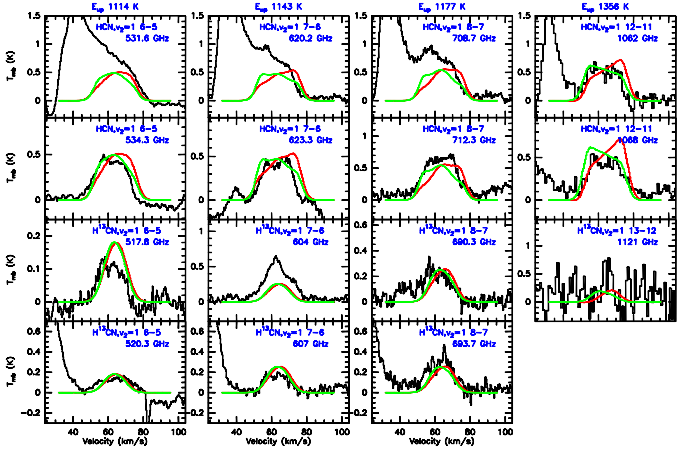}
  \caption{\textit{Herschel}/HIFI observations of vibrationally excited HCN in SgrB2(M), overlaid with the pure-infall (green) and the expansion+infall model (red). The kinematics are not well constrained in this hot, innermost region.}
  \label{fig:hifi_hcnvib}
\end{figure}

\textit{Herschel} is a new space observatory with a 3.5 m antenna operating at submm to far-infrared wavelengths \citep{pilbratt10}. Its spectral line receiver is HIFI \citep[Heterodyne Instrument for the Far Infrared,][]{degraauw10}. 
The guaranteed time key project HEXOS \citep[\textit{Herschel}/HIFI Observations of EXtraOrdinary Sources,][]{Bergin10} includes a full line survey of SgrB2(M) ($\alpha_{J2000} = 17^h47^m20.35^s$ and $\delta_{J2000} =
-28^{\circ}23'03.0''$). In March 2010, this was started with HIFI bands 1a, 1b, 2a, and 4b, providing coverage of the frequency ranges 479 $-$ 726 GHz and 1051 $-$ 1121 GHz. Each band contains one rotational transition from HCN, H$^{13}$CN, and HC$^{15}$N (Fig.~\ref{fig:hifi_hcn}), and two rotational transitions within their first vibrational state, which lies around 1000 K higher in energy (Fig.~\ref{fig:hifi_hcnvib}). Transition frequencies of HCN \citep{HCN_rot_2003,H13CN_HC15N_rot_2004} are taken from CDMS \citep{CDMS1_2001,CDMS2_2005}.
HIFI spectral scans are carried out in dual beam switch (DBS)
mode, where the DBS reference beams lie approximately 6$'$ 
apart. The wide band spectrometer (WBS) is used as a back-end,
providing a spectral resolution of 1.1 MHz over a 4-GHz-wide
intermediate frequency (IF) band. 
To allow separation of the two sidebands, bands 1a, 1b, 2a, and 4b were scanned
 with a redundancy of 4, 8, 8, and 4, respectively \citep{Comito02}.
The data were calibrated using the standard pipeline released with
version 2.9 of HIPE \citep{Ott10}, and subsequently exported to
CLASS\footnote{{\it Continuum and Line Analysis Single-dish Software},
  distributed with the GILDAS software, see 
http://www.iram.fr/IRAMFR/GILDAS.} using the HiClass task within HIPE. 
Deconvolution of
the DSB data into single-sideband (SSB) format was performed using
CLASS. All the HIFI data presented here, spectral features \emph{and}
continuum emission, are deconvolved SSB spectra. The horizontal and vertical polarizations differ by around 10\%, and were averaged. The intensity scale is main-beam temperature, and results from applying a beam efficiency
correction of about 0.68.

\begin{figure}
  \centering
  \includegraphics[angle=-90,width=0.49\textwidth]{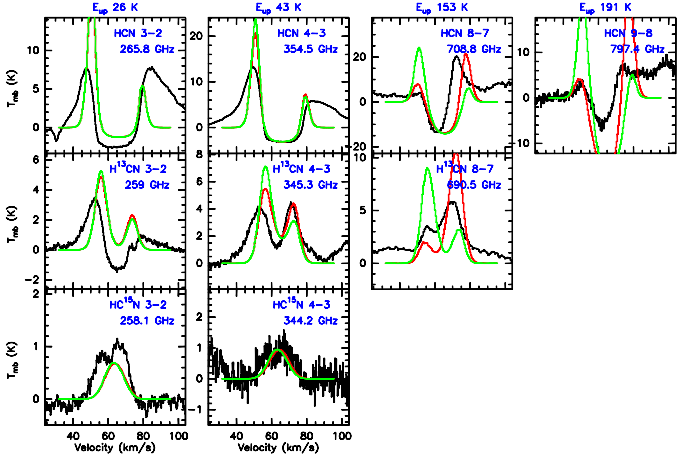}
  \caption{APEX observations of HCN in SgrB2(M), overlaid with the pure-infall (green) and the expansion+infall model (red). Although less clear than in the HIFI data, the same asymmetry change can be seen.}
  \label{fig:apex_hcn}
\end{figure}

APEX (Atacama Pathfinder Experiment) is a 12-m telescope located at 5100 m
altitude in the Atacama Desert, Chile, one of the optimal sites for submm observations on Earth \citep{Guesten06}. The lines presented here (Fig.~\ref{fig:apex_hcn}) are from a study of the structure of 12 hot cores (Rolffs et al. 2010, in prep.). Since the baseline levels in the ground-based data are not determined reliably enough to extract the continuum, the 850$\mu$m continuum is taken from ATLASGAL \citep{Schuller09}.

\begin{figure}
  \centering
  \includegraphics[angle=0,width=0.25\textwidth,bb=220 110 720 550]{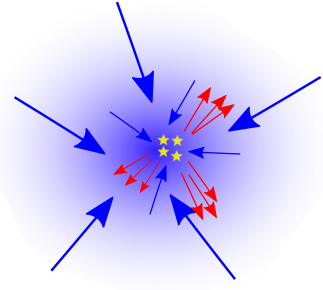}
  \caption{Sketch of the scenario described in the text.  In the outer parts of the cloud, infall dominates.  In the inner part, feedback has set in, in this example envisioned as multiple outflows, and gas is expelled from it. At the same time, infall continues along certain paths.}
  \label{fig:infall}
\end{figure}

\section{Modeling}

The spherical Monte Carlo radiative transfer code RATRAN \citep{Hogerheijde00} was employed to compute the molecular lines and the dust continuum. Using an iterative by-eye comparison, the fit to the data was obtained in two steps. First, the physical structure (density and temperature) was adapted to match the continuum. Second, the HCN abundance and the velocity were optimized for the lines.

The model we present here is centrally heated, its density following a radial power law with index 1.5. The dust opacity was taken from \citet{Ossenkopf94}, without either ice mantles or coagulation. 
At the inner radius, 1340 AU, the density is 1.8$\times$10$^8$ cm$^{-3}$ and the temperature is 1500 K. The temperature was computed in an approximate way from central heating, using the diffusion equation in the inner part and balance between heating and cooling in the outer part, assuming an effective temperature for the dust photosphere of 58 K. The temperature has a steeper gradient in the inner part, and falls off to 20 K at the outer radius of 5.6 pc, where the density is 7$\times$10$^3$ cm$^{-3}$. The model reproduces the continuum very well (Table~\ref{tab:cont}). Its central dust optical depth is 0.4 at 345~GHz, 0.8 at 500~GHz, 1.5 at 700~GHz, and 3.7 at 1.1~THz.

%

\begin{table}
\caption[]{Continuum fluxes of SgrB2(M) and the model. }
\label{tab:cont}
\renewcommand{\footnoterule}{}  
\begin{tabular}{|l c c c|}    
\hline\hline                 
Frequency & Beam size &  SgrB2(M)  & model \\
~[GHz] & [''] & [K] &  [K] \\
\hline
345 & 18.2 & 3.8 &  3.5 \\ 
\hline
500 & 43.1 & 2.4  & 2.4 \\ 
600 & 35.9 & 3.8  & 3.8 \\ 
700 & 30.8 & 5.7 & 5.7 \\ 
1100 & 19.6 & 14.6 & 14.2 \\ 
\hline
\end{tabular}
\\
{\bf Note:} The first row is the 850$\mu$m flux from LABOCA on APEX, which contains also lines. The next rows are the baseline levels at the given frequencies in HIFI bands 1a, 1b, 2a and 4b. The baseline level of HIFI is a reliable continuum measurement, since it is as strong as saturated absorption lines and compares well with ground-based continuum data.
\end{table}

\begin{figure}
  \centering
  \includegraphics[angle=-90,width=0.48\textwidth]{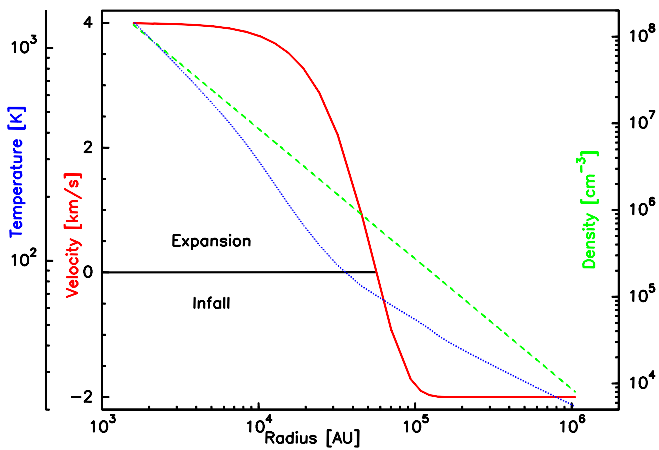}
  \caption{Physical structure of the model. Density is plotted as dashed green, temperature as dotted blue. The velocity field of the infall+expansion model is shown as the solid red curve (the infall model has a constant velocity of -2 km/s). At the transition between infall and expansion, the temperature is 65 K and the density is 6$\times10^5$ cm$^{-3}$.}
  \label{fig:model_vel}
\end{figure}

To compute the lines, we used a turbulent 1/e half width of 7 km/s  (12 km/s FWHM). The lines were additionally broadened by their high optical depth. We found no evidence of significant variations in the turbulent width with radius. The HCN abundance increases with temperature from 10$^{-8}$ at temperatures below 100 K, to 3$\times$10$^{-8}$ between 100 and 300 K, to 3$\times$10$^{-7}$ at temperatures above 300 K. The first jump is mainly needed to reproduce the intensities of the isotopologues and of vibrationally excited HCN, the second jump serves to fit vibrationally excited H$^{13}$CN.  These jumps could be due to evaporation of ice mantles and/or increased chemical production. We used a $^{12}$C/$^{13}$C ratio of 20 and $^{14}$N/$^{15}$N of 600 \citep{Wilson94}. The green model has a constant infall velocity of 2 km/s, and the red model has the same velocity in its outer parts, but an expansion in its inner parts. Figure~\ref{fig:model_vel} shows the velocity field.  We note that the exact shape of this, particularly the transition from infall to outflow, is not constrained by the observations, hence modeling the true velocity field by adopting this spherical approximation is an oversimplification (see Fig.~\ref{fig:infall}).

A comparison with the observations (Figs.~\ref{fig:hifi_hcn}, \ref{fig:hifi_hcnvib}, and \ref{fig:apex_hcn}) shows that the red model reproduces more closely the observed features. In particular, to reproduce the changes in the asymmetry that occur both from lower to higher J and from higher to lower optical depth (from HCN 6--5 to H$^{13}$CN 6--5), a change in the velocity field is required, from infall in the colder, outer parts to expansion in the warmer, inner parts.

\section{Discussion}

We now discuss the velocity field and how it is physically determined.

In a gravitationally contracting sphere with density power-law index $p$, a first guess and upper limit to the infall speed would correspond to free fall, i.e., ${\rm v}_{\rm ff}(R)=\sqrt{\frac{2GM(R)}{R}}$, where the mass inside the radius $R$ is $M(R) \propto R^{3-p} $ (for $p<3$), such that ${\rm v}_{\rm ff} \propto R^{\frac{2-p}{2}}$. In our model, this free-fall speed is roughly (at a radius of 1 pc) 10 times higher than the 2 km/s of the model. Such high infall velocities can be excluded because they would produce lines that are far more asymmetric than observed.

The mass accretion rate $\dot{M} = {\rm v}_{\rm acc}(R) \frac{{\rm d}M(R)}{{\rm d}R}$, so ${\rm v}_{\rm acc} \propto R^{p-2}$ for a constant mass accretion rate. 
Hence, infall accelerates in a spherical steady-state model with $p=1.5$. However, an accelerating infall, at least in the inner part, is inconsistent with the data. 
For simplicity, we chose a constant infall velocity, corresponding to accretion rates onto the cluster of 3$\times$10$^{-1}$ M$_\odot$/yr at the outer boundary and 10$^{-1}$ M$_\odot$/yr at a radius of 10$^{5}$ AU. The excretion rate reaches 8$\times$10$^{-2}$ M$_\odot$/yr at 2$\times$10$^{4}$ AU.

The real velocity field must be non-spherical, and infall could well continue in the inner part  (see Fig.~\ref{fig:infall} and also \citet{Qin08}). The peak on top of the HCN 12--11 absorption (Fig.~\ref{fig:hifi_hcn}) could be a hint of that infall. Since the H$^{13}$CN transitions are less asymmetric than modeled points in that direction, one may suspect that a linear combination of the pure infall and expanding models would produce the line spectra very well.  For the HCN lines, because of their high optical depths, such a linear combination of intensities does not naturally reproduce the observed spectra; more sophisticated modeling, in which we abandon the spherical approximation, would be required. For overlying model and data (Figs.~\ref{fig:hifi_hcn}, \ref{fig:hifi_hcnvib}, and \ref{fig:apex_hcn}), we assumed a source velocity of 64 km/s, although 60 km/s would reproduce more closely the high-J lines. This blue shift of absorption towards higher J is consistent with our picture of a velocity field changing from infall to expansion, and is indicative of a deviation from spherical symmetry.

While gravity is clearly responsible for the infall, there must also be feedback mechanisms from newly formed stars working against gravity, thus regulating star formation. The velocity of the infalling envelope must be lower than that of  pure gravitational collapse by around 90\%. Apart from magnetic braking, turbulence induced by bipolar outflows may be such a long-range feedback mechanism providing the necessary pressure \citep{Wang10}. The massive stars in the center dominate the luminosity, and drive the expansion by means of stellar winds, radiation pressure (including reprocessed light), and  heating and ionization of the gas, in particular thermal pressure of their H{\sc ii} regions \citep{Krumholz09}. Since in our model only 1\% of the total mass of 7$\times$$10^5$ M$_\odot$ is expanding, the outward momentum remains much lower than the inward momentum, indicating that feedback is not yet able to disrupt the cloud.

\section{Conclusions}

The changing asymmetry of the HCN lines clearly indicates a reversal of infall. Expansion motions dominate in the inner, warmer parts, and infall dominates in the outer, colder parts of the core. This can be naturally explained by the onset of feedback from massive stars. Our radiative transfer modeling demonstrates the power of constraining the source structure by fitting line shapes.  A model that successfully reproduces all the features cannot have spherical symmetry, and the exact geometry probably has to be constrained by interferometric observations, to determine the free parameters.  Although the \textit{Herschel}/HIFI data do not have the desired spatial resolution, they place strong constraints on any such model, since they have to be reproduced.  

We propose a scenario in which the cloud gravitationally contracts at significantly below the free-fall speed. A star cluster forms by fragmentation and accretion. The stars, especially the massive stars that dominate the luminosity, provide various feedback mechanisms to counteract the contraction. The additional pressure, be it thermal, radiative, or turbulent, decelerates the contraction, then dominates first in the inner part, before finally disrupting the whole cloud. SgrB2(M) is just beginning to drive out the gas, while large-scale global infall and probably also infall among very localized pathways in the interior remains ongoing.

This study illustrates the importance of high-frequency lines in constraining the source structure, and demonstrates the great potential of HIFI, which delivers velocity-resolved spectra at these frequencies. Ground-based telescopes may not be able to reach high enough frequencies. In our case, detecting the HCN 12--11 line was necessary in order to unambiguously trace the reversal of infall, so this result is unique to HIFI. We expect that a systematic study of high-J HCN transitions, even higher than the 12--11 line seen here, with a high signal-to-noise ratio, and toward a sample of hot core sources, would be highly rewarding in finding more sources in this particular stage of evolution.

\begin{acknowledgements}
  HIFI has been designed and built by a consortium of institutes and university departments from across 
Europe, Canada and the United States under the leadership of SRON Netherlands Institute for Space
Research, Groningen, The Netherlands and with major contributions from Germany, France and the US. 
Consortium members are: Canada: CSA, U.Waterloo; France: CESR, LAB, LERMA,  IRAM; Germany: 
KOSMA, MPIfR, MPS; Ireland, NUI Maynooth; Italy: ASI, IFSI-INAF, Osservatorio Astrofisico di Arcetri- 
INAF; Netherlands: SRON, TUD; Poland: CAMK, CBK; Spain: Observatorio Astron�mico Nacional (IGN), 
Centro de Astrobiolog�a (CSIC-INTA). Sweden:  Chalmers University of Technology - MC2, RSS \& GARD; 
Onsala Space Observatory; Swedish National Space Board, Stockholm University - Stockholm Observatory; 
Switzerland: ETH Zurich, FHNW; USA: Caltech, JPL, NHSC.
Support for this work was provided by NASA through an award issued by JPL/Caltech.
CSO is supported by the NSF, award AST-0540882.
\end{acknowledgements}


\end{document}